\documentclass{aa}
\usepackage{graphicx}
\usepackage{txfonts}

\newcommand{\gr}{$\gamma$-ray \,}

\begin{document}
   \title{Direct evidence of efficient cosmic ray acceleration and
     magnetic field amplification in Cassiopeia~A}

   \author{E. G. Berezhko
          \inst{1}
          \and
          H. J. V\"olk\inst{2}
          }

   \offprints{H. J. V\"olk}

   \institute{Institute of Cosmophysical Research and Aeronomy,
                     31 Lenin Ave., 677891 Yakutsk, Russia\\
             \email{berezhko@ikfia.ysn.ru}
         \and
             Max-Planck-Institut f\"ur Kernphysik,
                Postfach 103980, D-69029 Heidelberg, Germany\\
             \email{Heinrich.Voelk@mpi-hd.mpg.de}
             }

   \date{Received  / Accepted }

   \abstract{
It is shown that the nonlinear kinetic theory of
cosmic ray (CR)  acceleration in supernova remnants (SNRs)
describes the shell-type nonthermal X-ray morphology of Cas~A,
obtained in Chandra observations, in a satisfactory way. The set of
empirical parameters, like distance, source size and total energy release,
is the same which reproduces the dynamical properties of the SNR and the
spectral characteristics of the emission produced by CRs. The extremely
small spatial scales in the observed morphological structures at hard
X-ray energies are due to a large effective magnetic field
$B_\mathrm{d}\sim 500$~$\mu$G in the interior which is at the same time
not only required but also sufficient to 
achieve the excellent agreement between the spatially
integrated radio and X-ray synchrotron spectra and their calculated
form. The only reasonably thinkable condition for the production of such
a large effective field strength is a very efficiently accelerated nuclear
CR component. Therefore the Chandra data confirm first of all the
inference that Cas~A indeed accelerates nuclear CRs with the high
efficiency required for Cas~A to be considered as a member of the main
class of Galactic CR sources and, secondly, that the nonlinear kinetic
theory of CR acceleration in SNRs is a reliable method to determine the
magnetic field value in SNRs.
   }
   
  \authorrunning{Berezhko and V\"olk} 
  
  \titlerunning{Direct evidence of efficient CR acceleration in Cas~A}

   \maketitle
%

\section{Introduction}

Supernova remnants (SNRs) are the main sources of energy for the
Interstellar Medium (ISM). They control the physical state of the ISM, and
that presumably includes the nonthermal component of Interstellar matter,
often called the Galactic Cosmic Rays (CRs). 
Quantitative analysis of the SNR dynamics and its energetic particle
production requires
the determination of the
effective magnetic field in SNRs. For this we note that the magnetic field
decisively influences the acceleration of CRs and their subsequent
dynamics in SNRs. In the following we shall present evidence for the case
of Cassiopeia~A (Cas~A) that the nonlinear kinetic theory for CR
acceleration in SNRs (Berezhko et al. \cite{byk96}; 
Berezhko \& V\"olk \cite{bv97}; \cite{bv00})
does not only satisfactorily explain the observed SNR dynamics and the
properties of the nonthermal emission, but that it also
permits to determine the value of
the effective magnetic field in SNRs, as already demonstrated for
the case of SN~1006 (Berezhko et al. \cite{bkv02}; \cite{bkv03}).

The existence of a powerful population of relativistic particles in SNRs
is known from their nonthermal radiation. For some SNRs this emission is
detected in a very wide range from the radio to the \gr band. Cas~A
represents such an example. It is a shell-type SNR, and a bright
source of synchrotron radiation observed from radio waves (e.g. Baars et
al. \cite{baars}) to hard X-rays (Allen et al. \cite{all97}). 
These emissions are direct
evidence for the existence of a large number of relativistic electrons
presumably accelerated at the shock, with energies reaching about 10~TeV.  
Direct experimental knowledge of nuclear CR production can only be
obtained by high energy \gr measurements. If protons are efficiently
accelerated in Cas~A, then the $\pi^0$-decay $\gamma$-ray
spectrum as a result of inelastic collisions with the background nuclei
must extend beyond 1~TeV with a hard power-law spectrum. The detection of
a corresponding if
weak signal in TeV $\gamma$-rays has been indeed reported by the
HEGRA collaboration (Aharonian et al. \cite{aha01}).

According to the analysis by Borkowski et al. (\cite{bor96}) which is consistent
with the observed dynamics of Cas~A, the supernova shock expands into a
non-uniform circumstellar medium strongly modified by the intense
progenitor star wind. This circumstellar environment consists of a tenuous
inner bubble, created by the Wolf-Rayet wind, a dense shell of swept-up
slow Red Supergiant (RSG) wind material, and a subsequent unperturbed RSG
wind. The medium further out is not relevant here.  Since the shock has
already passed through the dense shell and has swept up a mass of about
5$M_{\odot}$ (Favata et al. \cite{favata}), a detectable $\pi^0$-decay \gr flux is
to be expected. Recently Chevalier \& Oishi (\cite{chevo}) have argued that
a Wolf-Rayet phase and
the corresponding modification of the preexisting RSG wind is not
required. To reach a final conclusion about this
important issue a detailed
dynamical and kinetic consideration has to be made, similar to
the one contained in our previous paper
(Berezhko et al. \cite{bpv03}). At the moment we can
only say that the global characteristics of nonthermal radiation of Cas~A
are expected to be not sensitive to these details of
the structure of circumstellar medium, because they are mainly determined
by the total amount of the swept up gas and the magnetic field
value in the remnant. Since the observed hard X-ray synchrotron emission
of Cas~A comes from the thin region near the current SN shock position,
the properties of this radiation are not expected to be
very sensitive to the previous history of the SNR evolution. Here we
use our model (Berezhko et al.  
\cite{bpv03})  which agrees with all previous data, leaving the question
about the role of Wolf-Rayet stage for future 
scrutiny, and test its consistency with the hard X-ray
morphology.

Comparing the nonthermal radio and X-ray synchrotron data with the
calculated spectrum of the energetic electrons, Berezhko et al.
(\cite{bpv03})  have inferred that the existing data require
very efficient acceleration of {\it CR nuclei} at the SNR blast wave which
has converted already several percent of the initial SNR energy content
into nuclear CR energy despite the necessity of renormalization of the
overall nuclear CR content due to the nonuniform injection of suprathermal
ions (V\"olk et al. \cite{vbk03}). At the same time a large interior magnetic
field strength $B_\mathrm{d} \approx 500 \mu$G is required. The large value of
$B_\mathrm{d}$ is expected and possible in this case, using the spatially
integrated radio and X-ray synchrotron spectra to empirically determine
the effective magnetic field strength and the injection rate in the theory
(see also V\"olk \cite{voel03}).

In fact there are very recent measurements of details of the X-ray
morphology of Cas~A with the Chandra telescope by Vink \&
Laming (\cite{vl03}).  These authors have used the detailed Chandra morphology to
estimate the magnetic field strength by assuming the small spatial X-ray
scale to be determined by postshock synchrotron losses. Our aim is to show
that the actual B-field strength coincides with that inferred from the
radio spectrum and nonlinear acceleration theory. We confirm that the
field is indeed amplified in Cas~A, although it turns out that the actual
value is significantly larger than estimated by Vink \& Laming (\cite{vl03}).

\section{Results and discussion}

Compared with SN~1006 which exploded
into an essentially uniform ISM, Cas~A is a very different and much more
complex object. The dynamics of the SN shock and the major fraction of the
emission produced by accelerated CRs are predominantly determined by the
parameters of the swept-up RSG shell. They depend less sensitively on the
parameters of the medium near the current shock position since that
presumably consists of free RSG wind material and contains little
mass. Nevertheless, since the required magnetic field in the interior
downstream of the shock is so large that it strongly influences the
energy spectrum of the CR
electrons as a result of their synchrotron losses (Berezhko
et al. \cite{bpv03}), the hard X-ray synchrotron emission should
come predominantly from the relatively thin postshock region, because
deeper inside the remnant the upper cutoff energy of CR electrons becomes
too low. Therefore one would expect that the nonlinear kinetic
theory should not only reproduce the spectral
characteristics of Cas~A in a satisfactory way, but should
also be able to explain the observed fine structure of the nonthermal
emission in hard X-rays.

For the comparison with the Chandra data we shall
not present the details of our model. They have already been described in
the above paper (Berezhko et al. \cite{bpv03}). In order to give a
clear interpretation and simple estimate of the
spatial variation of the synchrotron emission near the SNR shock 
we shall use some
simple analytical approximations,
extending the discussion of SN~1006 (Berezhko et al.  
\cite{bkv03}). As a second step we shall make a comparison of the Chandra
data with the results of the nonlinear theory.

It was shown (Berezhko et al. \cite{bkv03}) that due to the
strong synchrotron losses the very energetic electrons occupy
only a thin region behind the shock of thickness 
\begin{equation}
l_2^{-1}=(2l_\mathrm{d2})^{-1}(\sqrt{1+4l_\mathrm{d2}/l_\mathrm{c2}}-1),
\label{eq1}
\end{equation}
where $l_\mathrm{d}(p)=\kappa(p)/u$ and $l_\mathrm{c}(p)=\tau(p) u$ are the 
diffusive and convective
length scales respectively,
$\tau(p)= 9 m_\mathrm{e}^2c^2/(4 r_0^2B^2p)$,
is the synchrotron loss time, $m_\mathrm{e}$ is the electron mass, $r_0$
denotes the classical electron radius, $B$ is the effective magnetic
field strength, $\kappa(p)=\rho_\mathrm{B} v/3$ is CR diffusion coefficient, and
$\rho_\mathrm{B}(p)$, $p$ and $v$ are the particle gyroradius, momentum and
velocity, respectively. In addition $u_2=V_\mathrm{s}/\sigma$ is the
downstream plasma speed relative to the shock
front, $V_\mathrm{s}$ is the shock speed,  and $\sigma$ is the total shock
compression ratio.

Depending upon the ratio $l_\mathrm{d}/l_\mathrm{c}$ 
we have two different extreme cases.
For electron momenta $p$ corresponding to $l_\mathrm{c}\gg 4l_\mathrm{d}$ (weak
losses) we have $l_2=l_\mathrm{c}$. In this case the loss time $\tau$ 
is much larger than the acceleration time
$\tau_\mathrm{acc}\sim \kappa/u^2$ and therefore the losses do not influence the
electron spectrum during their acceleration. The electron spectrum has a
power law character up to a cutoff momentum 
$p_\mathrm{max}>p$,
which exceeds the considered momentum $p$. Later on, the losses in the
downstream region lead to a decrease of the cutoff momentum so that at the
distance $r=R_\mathrm{s}-l_\mathrm{c}$ it drops to the value
$p_\mathrm{max}(r)=p$. Here $R_\mathrm{s}$ is
the shock radius.
In the opposite extreme case of strong losses
$l_\mathrm{c}\ll 4l_\mathrm{d}$, we have 
$l_2=\sqrt{\kappa_2 \tau_2}$.
which is in fact independent of momentum if particle diffusion
proceeds in the Bohm limit. In this case losses become significant
already during the electron acceleration and therefore the cutoff momentum
$p_\mathrm{max}$ is lower than $p$, so that the corresponding electrons 
belong to the exponential tail of the
distribution. Since the nonthermal X-ray emission of Cas~A belongs to the
steep tail of the synchrotron spectrum, Eq.(\ref{eq1}) suggests that $l_\mathrm{c} \ll
l_\mathrm{d}$ for the radiating electrons.

Since the synchrotron data are given in terms of the emission
frequency $\nu$, it is useful to rewrite Eq.(\ref{eq1}), taking into account
the approximate relation $\nu\propto Bp^2$:
\begin{equation}
l_2=\sqrt{\kappa_2 \tau_2}/(\sqrt{1+\delta^2}-\delta),
\label{eq2}
\end{equation}
where 
\begin{equation}
\delta^2=l_\mathrm{c2}/(4l_\mathrm{d2})=0.12[c/(r_0\nu)][V_\mathrm{s}/(\sigma c)]^2.
\label{eq3}
\end{equation}

We then have the situation that the synchrotron emissivity
$q_\mathrm{\nu}(\epsilon_\mathrm{\nu},r)$ is everywhere close to zero except for a
thin radial region of thickness $\Delta r\sim l_2$ just behind the SN
shock, since the emissivity from the upstream region $r>R_\mathrm{s}$ is
insignificant.

In projection along the line of sight, the radial emissivity profile
determines the remnant's surface brightness $J_\mathrm{\nu}$. For the X-ray
energy $\epsilon_\mathrm{\nu}$ it has the form
\begin{equation}
J_\mathrm{\nu}(\epsilon_\mathrm{\nu},\rho)=
2\int_a^0 dx q_\mathrm{\nu}(\epsilon_\mathrm{\nu},r=\sqrt{\rho^2+x^2}), 
\label{eq4}
\end{equation}
where $\rho$ is the distance between the center of the remnant and the
line of sight. In the case of a spherical shock
$a=-\sqrt{R_\mathrm{s}^2-\rho^2}$.

To estimate the shape of the expected profile $J_\mathrm{\nu}(\rho)$
one can consider the simple case, when the
emissivity has a form 
\begin{equation}
q_\mathrm{\nu}(\epsilon_\mathrm{\nu},r)=q_2(\epsilon_\mathrm{\nu}) \exp[(r-R_\mathrm{s})/l_2]
\label{eq5}
\end{equation}
for $r\le R_\mathrm{s}$. In the case of interest, $l_2\ll R_\mathrm{s}$, the
main contribution to the integral (\ref{eq4}) comes from the thin regions
near the boundary $x=a$. To perform the integration along the line of
sight one can therefore expand the quantity $r=\sqrt{\rho^2+x^2}$ in
Eq.(\ref{eq5}) in powers
of $x-a$ up to the second order term which results in:
\[
J_\mathrm{\nu}=\frac{2q_2R_\mathrm{s}l_2}{\sqrt{R_\mathrm{s}^2-\rho^2}}
\left\{
1-
\frac{R_\mathrm{s}^2-2\rho^2}{R_\mathrm{s}^2z}\right.
-
\]
\begin{equation}
\hspace{0.5cm}
\left. 
-e^{-z}
\left[
1-
\frac{R_\mathrm{s}^2-2\rho^2}{R_\mathrm{s}^2z}
+\frac{(z+2)}{2}
\left(\frac{2\rho^2-R_\mathrm{s}^2}{R_\mathrm{s}^2}\right)
\right]
\right\},
\label{eq6}
\end{equation}
where $z=(R_\mathrm{s}^2-\rho^2)/(R_\mathrm{s}l_2)$.
This approximate expression is accurate to about
one percent for the case of a small thickness of the
emission layer $l_2\ll 0.1R_\mathrm{s}$. As it is clear from this expression the
distant observer will see the profile $J_\mathrm{\nu}(\rho)$ which, going from
large distances to the center of the SNR, starts at $\rho=R_\mathrm{s}$, reaches
its peak value 
$J_\mathrm{max}=2q_2\sqrt{R_\mathrm{s}^2-\rho_m^2}$ at $\rho_\mathrm{m}\approx
R_\mathrm{s}-0.85l_2$ and smoothly drops to $J_\mathrm{min}=2l_2q_2$ at the SNR center
$\rho=0$. Therefore the outer scale $L_1\approx l_2/2$
characterizes the emission profile at $\rho>\rho_\mathrm{m}$, whereas for the
inner scale, which characterizes the profile at $\rho<\rho_\mathrm{m}$, we have
$L_2\approx 13 L_1$. This means that the width $L=L_1+L_2\approx 7l_2$ of
the observed brightness profile $J_\mathrm{\nu}(\rho)$ is always appreciably
wider than the width $l_2$ of the emissivity layer $q_\mathrm{\nu}(r)$, simply
due to the geometry of the system.

According to (\ref{eq2}) 
the downstream magnetic field $B_\mathrm{d}$ is
\begin{equation}
B_\mathrm{d}=[3m_e^2c^4/(4er_0^2l_2^2)]^{1/3}(\sqrt{1+\delta^2}-\delta)^{-2/3}.
\label{eq7}
\end{equation}
The case of strong losses $l_\mathrm{c}\ll 4l_\mathrm{d}$ corresponds to $\delta=0$.
According to Vink \& Laming (\cite{vl03}) the total thickness of the observed
profile at $\epsilon_{\nu}>2.7$~keV is as small as $1.5''$. However,
following the above consideration, only the small fraction
$1.5''/7\approx 0.22''$ corresponds to the thickness of the downstream
emission region
$l_2$. For a distance $d=3.4$~kpc and an angular radius of $2.6'$ for 
Cas~A
this gives $l_2\approx 1.1\times 10^{16}$~cm. In the case of strong losses
$(\delta=0)$ 
$B_\mathrm{d}\approx 470$~$\mu$G, independently of synchrotron energy and
shock speed. 
For  $V\mathrm{s} = 3000$~km/s and $\sigma=6$, corresponding to the overall
nonlinear
model (Berezhko et al. \cite{bpv03}), $\delta^2=0.054$ resulting in 
$B_\mathrm{d}\approx 550$~$\mu$G. For a larger shock speed $V_\mathrm{s} = 5000$~km/s,
as derived from X-ray measurements and used by Vink \& Laming,
we have $\delta^2= 0.15$ and $B_\mathrm{d}\approx 610$~$\mu$G.

The corresponding magnetic field value is by a factor of about five larger
than estimated by Vink \& Laming (\cite{vl03}). There are two reasons for such a
large discrepancy (i) only a small part of
the observed profile width $L$ corresponds to the thickness $l_2$
of the emission region
(ii) the correct formula (\ref{eq1}), 
cf. V\"olk et al. (\cite{vmf81}), implies $l_\mathrm{c2}<l_2$. Using 
$l_\mathrm{c2}=L$, as was done by Vink \& Laming, therefore doubly
underestimates $B_\mathrm{d}$, since $l_\mathrm{c2}\propto B_\mathrm{d}^{-2}$.

\begin{figure}
\centering
\includegraphics[width=7.5cm]{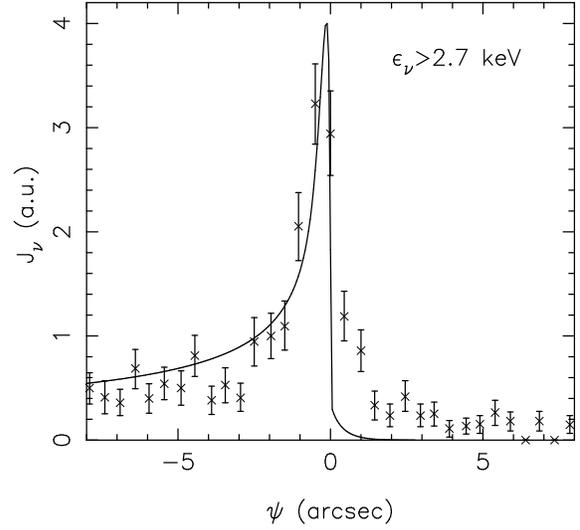}
\caption{The X-ray emissivity for
X-ray energies $2.7<\epsilon_\mathrm{\nu}<9$~keV as a function of angular  
radial distance. 
The Chandra 
measurements are shown (Vink \& Laming \cite{vl03}).}
\label{f1}
\end{figure}

The numerically calculated brightness profile
$J_\mathrm{\nu}(\rho)=\int_{\epsilon_1}^{\infty}d\epsilon_\mathrm{\nu} 
J_\mathrm{\nu}(\epsilon_\mathrm{\nu},\rho)$
for the X-ray energy
$\epsilon_\mathrm{\nu}>\epsilon_1=2.7$~keV is shown in Fig.\ref{f1}.  The sharpest
experimental X-ray brightness profiles obtained by the Chandra observers
(Vink \& Laming \cite{vl03}) are shown as a function of angular distance $\psi =2.6\times
60(\rho-R_\mathrm{s})/R_\mathrm{s}$ in arcseconds, taking into account that the angular
radial size of Cas~A is $2.6'$. Since the absolute values of the
measurements are not known, the theoretical profile is
normalized so as to obtain the best fit to the experimental profile. 
The very sharp part on the right side of
the calculated profile $J_\mathrm{\nu}(\psi)$ corresponds to the shock front
position $\rho=R_\mathrm{s}$. Compared with Vink \& Laming (\cite{vl03}) all experimental
points are shifted by $\Delta \psi =0.3''$ to the left. One can see that
the experimental values agree quite well with our calculations, which yielded
a postshock value of the magnetic field strength at the current
evolutionary epoch (see Berezhko et al. \cite{bpv03}) of
$B_\mathrm{d}=480$~$\mu$G.
%
Taking into account that the Chandra angular resolution is $0.5''$
we conclude that the actual downstream magnetic field in Cas~A is
indeed $B_\mathrm{d}\approx 500$~$\mu$G with an uncertainty of about 30\% which 
encompasses the above simple estimates of the local field.
  
There are a number of reasons for a broadening of the observed profile,
compared with the ideal case of a spherical shell. First of all, it is
clear from the observations that the actual shock front deviates from a
strictly spherical form. The wavy shape of the shock front can be produced
as the result of small scale density inhomogeneities of the ambient ISM.
Any small-scale distortion of the spherical emission shell leads to a
broadening of the observed brightness profile. This explains why the
observed X-ray radial profile is slightly broader than the theoretical
profile (see Fig.1). Therefore it is hard to believe that the
actual magnetic field in Cas~A could be significantly lower than the above
value.

Since at the current epoch the supernova shock propagates through the free
RSG wind, where the required effective magnetic field would be so strong
that the Alfv\'en speed is $c_\mathrm{a}\approx 160$~km/s (Berezhko et al.
\cite{bpv03}), this field exceeds the preexisting wind field by at least
an order of magnitude, because the wind should presumably be
superalfvenic, that is $c_\mathrm{a}<V_\mathrm{w}$, 
where $V_\mathrm{w}\sim 10$~km/s is the RSG wind
velocity. From this argument, the required
strong downstream magnetic field is inevitably the result of considerable
amplification near the supernova shock. As argued for the
case of SN~1006 (Berezhko et al. \cite{bkv03}) such a strong magnetic
field amplification can only be produced nonlinearly by a very efficiently
accelerated {\it nuclear CR component}. In this case their number,
consistent with all existing data, is so high that they are able to
strongly excite magnetohydrodynamic waves, and thus to amplify the
magnetic field, and at the same time to permit efficient CR scattering on
all scales, approaching the Bohm limit (Bell \& Lucek \cite{bluc01}).
The same large effective
magnetic field is required by the comparison of our selfconsistent theory
with the synchrotron observations.

For such a high downstream magnetic field $B_\mathrm{d}\approx 0.5$~mG, which
leads to strong synchrotron losses of the CR electron component, the
leptonic contribution to the TeV \gr emission is so small that it is far
below the detected HEGRA flux (Berezhko et al. \cite{bpv03}).

\section{Summary}

We conclude that the Chandra data confirm the magnetic field
amplification and the very efficient acceleration of nuclear CRs in
Cas~A, predicted by Berezhko et al. (\cite{bpv03}). This efficiency is
consistent with the requirements for the Galactic CR energy budget. At
the same time, together with SN~1006 
(Berezhko et al. \cite{bkv02}; \cite{bkv03}) this
is now a second -- and astrophysically very different -- case, where the
prediction of the nonlinear kinetic model for CR acceleration in SNRs is
fully confirmed experimentally. Therefore we conclude that this theory is
not only able to describe the CR dynamics and acceleration in SNRs, but
that it constitutes in addition a reliable method to quantitatively
determine the effective magnetic field strength which is produced in the 
acceleration process.

Since our model uses the Bohm diffusion of CRs, which is consistent with
the field amplification picture, one can consider the consistency of the
magnetic field values required to explain the global properties of the
synchrotron emission on the one hand (Berezhko et al. \cite{bpv03}), and to
reproduce the fine structure of the X-ray radiation on the other as
observational support for the Bohm limit of CR diffusion near the strong
SNR shock.

The authors thank L.T.Ksenofontov for his assistance in the preparation of
this paper. This work has been supported in part by the Russian Foundation
for Basic Research (grant 03-02-16524). EGB acknowledge the hospitality of
the Max-Planck-Institut f\"ur Kernphysik, where this work
was carried out.


\begin{thebibliography}{99}

\bibitem[2001]{aha01}
Aharonian, F.A., Akhperjanian, A., Barrio, J., et al. 2001, A\&A, 370,
112

\bibitem[1997]{all97}
Allen, G.E., Keohane, J.W., Gotthelf, E.V. et al. 1997, ApJ, 487, L97

\bibitem[1977]{baars}
Baars, J.W.M., Genzel, R, Paulini-Toth, I.I.K. \& Witzel, A. 1977,
A\&A, 61, 99

\bibitem[2001]{bluc01}
Bell, A. R. \& Lucek, S. G. 2001, MNRAS, 327, 433

\bibitem[1996]{byk96} 
Berezhko, E. G., Elshin V. K. \& Ksenofontov, L. T. 1996, JETP, 82, 1

\bibitem[1997]{bv97}
Berezhko, E. G. \& V\"olk, H. J. 1997, Astropart. Phys., 7, 183

\bibitem[2000]{bv00}
Berezhko, E.G. \& V\"olk, H.J. 2000, A\&A, 357, 183

\bibitem[2002]{bkv02}
Berezhko, E.\ G., Ksenofontov, L.\ T., V\"olk, H.\ J.\ 2002, A\&A,
395, 943

\bibitem[2003a]{bkv03}
Berezhko, E.\ G., Ksenofontov, L.\ T., V\"olk, H.\ J.\ 2003a, A\&A,
412, L11

\bibitem[2003b]{bpv03}
Berezhko, E.\ G., P\"uhlhofer, G., V\"olk, H.\ J.\ 2003b, A\&A,
400, 971

\bibitem[1996]{bor96}
Borkowski, K.J., Szymkowiak, A.E., Blondin, J.M. \& Sarazin, C.L. 1996,
ApJ, 466, 866

\bibitem[2003]{chevo}
Chevalier, R.A. \& Oishi, J. 2003. ApJ, 593, L23

\bibitem[1997]{favata}
Favata, F., Vink, J., Dal Fiume, D. et al. 1997, A\&A, 324, L49

\bibitem[2003]{vl03}
Vink, J. \& Laming, J.M. 2003, ApJ, 548, 758

\bibitem[2003]{voel03}
V\"olk, H.J. 2003, to appear in: Proc. 28th ICRC 
(Tsukuba), Plenary Talks; astro-ph/0312585. 

\bibitem[1981]{vmf81}
V\"olk, H.J., Morfill, G.E. \& Forman, M.A. 1981, ApJ, 249, 161

\bibitem[2003]{vbk03}
V\"olk, H.J., Berezhko, E.G. \& Ksenofontov, L.T. 2003, A\&A, 409,
563

\end{thebibliography}
\end{document}